\documentclass{article}
\usepackage{graphicx}
\usepackage{subfigure}
\usepackage{amsmath}
\usepackage{booktabs}
\usepackage{amssymb}
\begin{document}
\title{Contrastive Psudo-supervised Classification 
for Intra-Pulse Modulation of Radar Emitter Signals 
Using data augmentation}
\maketitle


\author{HanCongFeng, XinHaiYan,KaiLiJiang,XinYuZhao, BinTang}




\abstract{The automatic classification of radar waveform is a fundamental technique in electronic countermeasures (ECM).Recent supervised deep learning-based methods have achieved great success in a such classification task.However, those methods require enough labeled samples to work properly and in many circumstances, it is not available.To tackle this problem, in this paper, we propose a three-stages deep radar waveform clustering(DRSC) technique to automatically group the received signal samples without labels.Firstly, a pretext model is trained in a self-supervised way with the help of several data augmentation techniques to extract the class-dependent features.Next,the pseudo-supervised contrastive training is involved to further promote the separation between the extracted class-dependent features.And finally, the unsupervised problem is converted to a semi-supervised classification problem via pseudo label generation. The simulation results show that the proposed algorithm can effectively  extract class-dependent features, outperforming several unsupervised clustering methods, even reaching performance on par with the supervised deep learning-based methods.}


\newcommand\keywords[1]{\textbf{Keywords}: #1}

%


\section{Introduction}
Automatic classification of radar waveform is an essential task for modern radar countermeasures. The accurate classification of the radar waveform could help the radar reconnaissance systems perform pulse deinterleaving and determine the function and type of enemy radars.
With the rapid development of radar systems\cite{bluntOverviewRadarWaveform2016}, new types of radars waveform continue to emerge, which brings forward the higher request for the radar reconnaissance systems.

At present, researchers mainly focus on deep learning(DL) based methods to classify radars waveforms.Some DL-based methods convert the radar signals into time‐frequency images (TFIs) and utilize DL network architectures in the fieldof computer vision to classify TFIs. Wang et.al\cite{wangAutomaticRadarWaveform2017} uses Wigner-Ville distribution to generate TFIs and design a CNN to perform image classification that outperforms the auto-correlation-based methods. In \cite{quRadarSignalIntrapulse2019}, the convolutional denoising autoencoder was proposed to classify 12 intra-pulse modulation signals. This method can remove the noise of the TFIs and achieve higher recognition accuracy under low signal-to-noise ratio (SNR).Thien et.al\cite{huynh-theAccurateLPIRadar2021}  designs a multi-scale CNN with skip-connection to recognize the low probability of intercept (LPI) radar waveforms using  Choi-Williams TFIs under channel impairment.This method significantly outperforms previously proposed deep models.Other researchers design the network structure using raw one-dimensional(or two-dimensional IQ ) radar signals as input such as the CNN and its variants\cite{wuRadarEmitterSignal2020,yuanIntraPulseModulationClassification2021}, the long short time memory(LSTM) network-based methods\cite{weiIntrapulseModulationRadar2020}.Without using any feature extraction technique, those methods have comparable performance as the methods using the TFIs.
Those DL-based methods are supervised which require a vast amount of labeled signals to work properly.However,in practical scenarios of warfare,the intercepted waveforms may always contain unknown modulations, although the model can be re-trained to adapt the new environment, labeling the data can be very time-consuming and require requires an enormous amount of expert knowledge\cite{SignalEyeAISoftware,yuanSemiSupervisedClassificationIntraPulse2022}.To tackle the problem of unknown recognition, Liu\cite{liuUnsupervisedRadarSignal2022} proposed a Multi‐block Multi‐view Low‐Rank SparseSubspace-based Clustering method to cluster the TFIs of 12 radar signal types.However, there is still a performance gap compared with the supervised methods.

Recent advances in self-supervised representative learning have made great success in the field of Natural Language Processing(NLP) and computer vision(CV).Devlin et.al\cite{devlinBERTPretrainingDeep2019} propose a bidirectional transformer to predict the Masked tokens in a self-supervised manner, that outperforms previous state-of-the-art  methods by a large margin in 11 downstream tasks.Chen et.al\cite{chenSimpleFrameworkContrastive2020} proposed a simple framework for contrastive learning of visual representations(SimCLR), which outperforms sprevious supervised models on ImageNet use only  1$\%$ labeled data.Based on that, several image clustering algorithms\cite{niuSPICESemanticPseudolabeling2022,vangansbekeSCANLearningClassify2020} were proposed and achieved performance close to the supervised algorithm on datasets like CIFAR-10 and STL-10.

The success of those self-supervised methods  heavily relies on the choice of the data augmentation policy. For example, in SimCLR, the performance can have up to 30 $\%$differnce when different augmentation was employed.The direct use of some augmentation policy for image classification on TFIs may be inappropriate, for example,the rotation of the TFI of a single-carrier frequency (SCF) signal could turn it into a linear frequency modulation (LFM) signal.We believe an appropriate  data augmentation policy should consider the signal model of the radar signals, which is usually represented by  complex-valued operations. Therefore, to simplify the augmentation policy and reduce the computational costs, instead  of converting  to TFIs, the raw IQ signals are used as the input.
The main contributions of this paper are listed as follows:
\begin{enumerate}
	\item	Propose a 3-stage  unsupervised deep clustering algorithm that can achieve performance close to the supervised counterparts. 
	\item Propose a series of data augmentation methods that improve the performance of both supervised and unsupervised training.
	\item Design a network architecture  that significantly improves the performance of unsupervised clustering.

\end{enumerate}
The rest of the paper is organized as follows: Section \ref{Methods} introduces our three-stages unsupervised deep radar signal clustering algorithm together with the proposed network architecture .Section \ref{Results} describes the setup of the experiment, and the comparison with  the state-of-the-art. Section \ref{Discussion} analyses the output features and discuss a practical application a scenario of the proposed method.And finally, section \ref{Conclusions} summarizes the whole paper.

\section{ The proposed method}
\label{Methods}
\begin{figure}[htbp]
	\centering
	\includegraphics[width=0.8\textwidth]{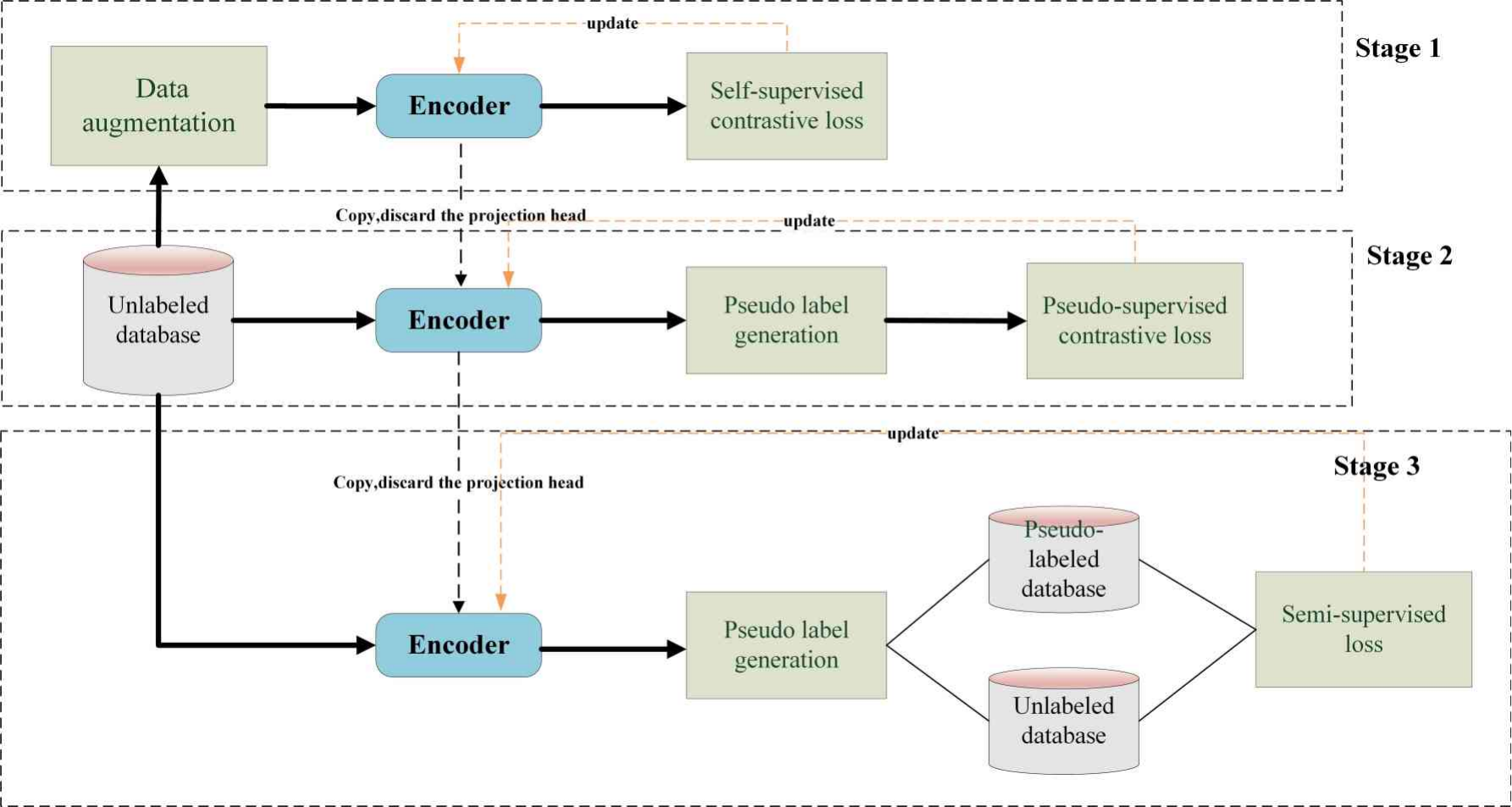}
	\caption{The three stages of the proposed method}
	\label{method}
\end{figure}

The gap between supervised and unsupervised learning is mainly caused by the entanglement of class-dependent and class-independent features. Therefore, the proposed method boosts unsupervised learning by removing the class-independent features and using pseudo-supervision. It consists of three stages, which are the contrastive pretext training stage, the pseudo-supervised contrastive training stage and the semi-supervised self-labeling stage,as shown in Figure \ref{method}.The following subsections present a detailed description of each stage of the proposed method.

\subsection{The contrastive pretext training}
The extraction of stable characteristics that only depend on the signal class serves as the first step in the signal classification process. Different transform domain analyses \cite{ravikishoreAutomaticIntrapulseModulation2017} are used in traditional radar signal processing approaches to accomplish this. Unfortunately, none of these can be suitable for all types, and it is unfeasible to develop a transformation manually for unidentified modulation.As with many supervised DL-based methods, an effective feature extraction approach needs to be data-driven. To extract class-dependent features, the network is guided by ground-truth labels in supervised training methods, which is not possible in unsupervised learning.
To address this issue, a deep neural network is designed to be a no-linear filter to remove the class-independent features while maintaining sample discrimination, where the filtered features should be class-dependent. 
\subsubsection{Self-supervised contrastive training}
Contrastive training is suggested to perform instance discrimination to ensure consistency across radar signals and the augmentations, similar to  SimCLR\cite{chenSimpleFrameworkContrastive2020}. Formally, the model is optimized to minimize the objective function as follows:
\begin{align}
	\label{self-supervisedcontrastiveloss}
	{{\cal L}^{self}} = - \sum\limits_{i = 1}^{2N} {\log \frac{{\exp \left( {{{\rm{e}}_{\rm{i}}} \cdot {{\rm{e}}_{{\rm{j}}({\rm{i}})}}/\tau } \right)}}{{\sum\limits_{{\rm{k}} = 1}^{2{\rm{N}}} {{{\rm{l}}_{[{\rm{k}} \ne {\rm{i}}]}}}  \cdot \exp \left( {{{\rm{e}}_{\rm{i}}} \cdot {{\rm{e}}_{\rm{k}}}/\tau } \right)}}}
\end{align}
,where ${{\rm{e}}_{\rm{i}}}={{{\Phi }_\theta }\left( {{X_i}} \right)}$,${{\rm{e}}_{{\rm{j}}({\rm{i}})}}={{{\Phi }_\theta }\left( {Arg\left[ {{X_i}} \right]} \right)}$ represent the embedding of the $ith$ sample and its augmented version, and the term ${\Phi }_\theta$ represents an encoder neural network with parameter $\theta$. $\tau$ is the temperature parameter and $N$ is the size of one minibatch. By minimizing the contrastive loss, the negative examples of the anchors will be uniformly distributed on the embedding hypersphere while similar samples  will be mapped together. When the augmentation $Arg\left[{{\cdot}}\right]$ only changes some class-independent properties of the signal, the encoder neural network is forced to discard them while keeping between-class  discrimination to minimize Eq \ref{self-supervisedcontrastiveloss}.Therefore,the features extracted by the encoder have to be class-dependent.
\subsubsection{Data augmentation for radar signals}
\begin{figure}[htbp]
	\centering
	\subfigure[]{
		\includegraphics[width=0.3\textwidth]{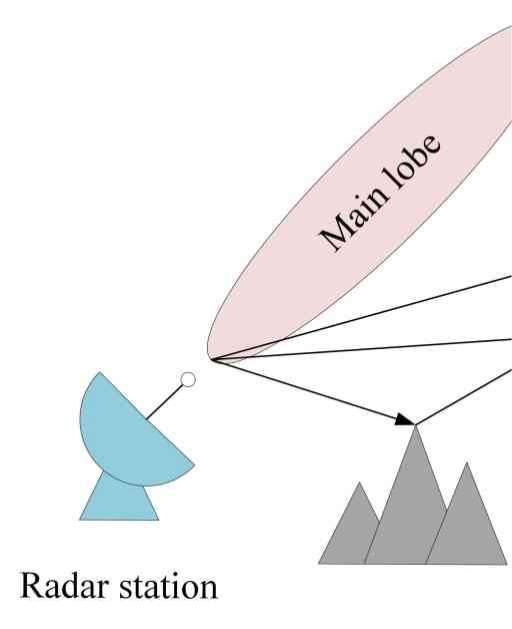}}
	\subfigure[]{
		\includegraphics[width=0.3\textwidth]{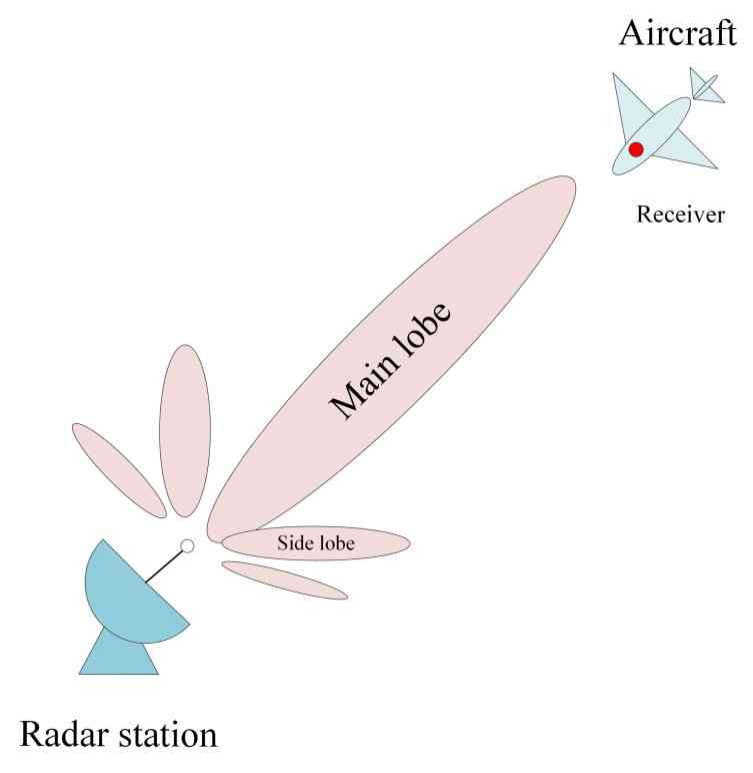}}
	\caption{Different scenarios of radar signals receiving: (a)receiving through multipath, (b)receiving through free-space}
	\label{schematic}
\end{figure}

The key of designing $Arg\left[{{\cdot}}\right]$ is to identify the class-independent features in the received signal samples.Thus, the signal model for the radar signal needs to be investigated.Refer to\cite{huynh-theAccurateLPIRadar2021}, the received discrete signal can be modeled as
\begin{align}
	\label{signal_model}
	y(k)=\exp \left(j \cdot\left(\theta(k)+2 \pi \frac{f_{c}}{f_s} k\right)+\theta_{p}\right) \otimes h(k)+n(k)
\end{align}
where $x(j\cdot\theta(k))$ represent the baseband noise-free radar signal with instantaneous phase function $\theta(k)$, $f_c$ is the carrier frequency,$f_s$ is the sampling rate.$\theta_{p}$ is the phase delay due to the propagation. $h(k)$ is the impulse response of the channel and $n(k)$ represents the additive white Gaussian noise.
In the real world, some radar signals may be received in a free-space environment
while others may pass through a rayleigh fading channel due to the multi-path effect. For example, when a phase array radar is tracking an airplane, the radar warning receiver(RWR) is likely to receive signals from only one path, due to the strong directionality of the phase array antenna. While in other cases, like the radar is searching a target in the valley, the signals may be reflected by the obstacles before receiving.A schematic for two cases is shown in Figure \ref{schematic}.Therefore, $h(k)$ can be expressed as 

\begin{align}
	h(k) & = \left\{
	\begin{aligned}
	A\delta(k-k_0) ,\quad p & = p_0\\
	Rayleigh(k), \quad   p & = 1-p_0
	\end{aligned}
	\right.
	\label{channel}
\end{align}
where $k_0$ is the path delay,$A$ is the path attenuation and $p_0$ is the probability of receiving from free-space propagation.

\begin{figure}[htbp]
	\centering
	\subfigure[original signal]{
		\includegraphics[width=0.3\textwidth]{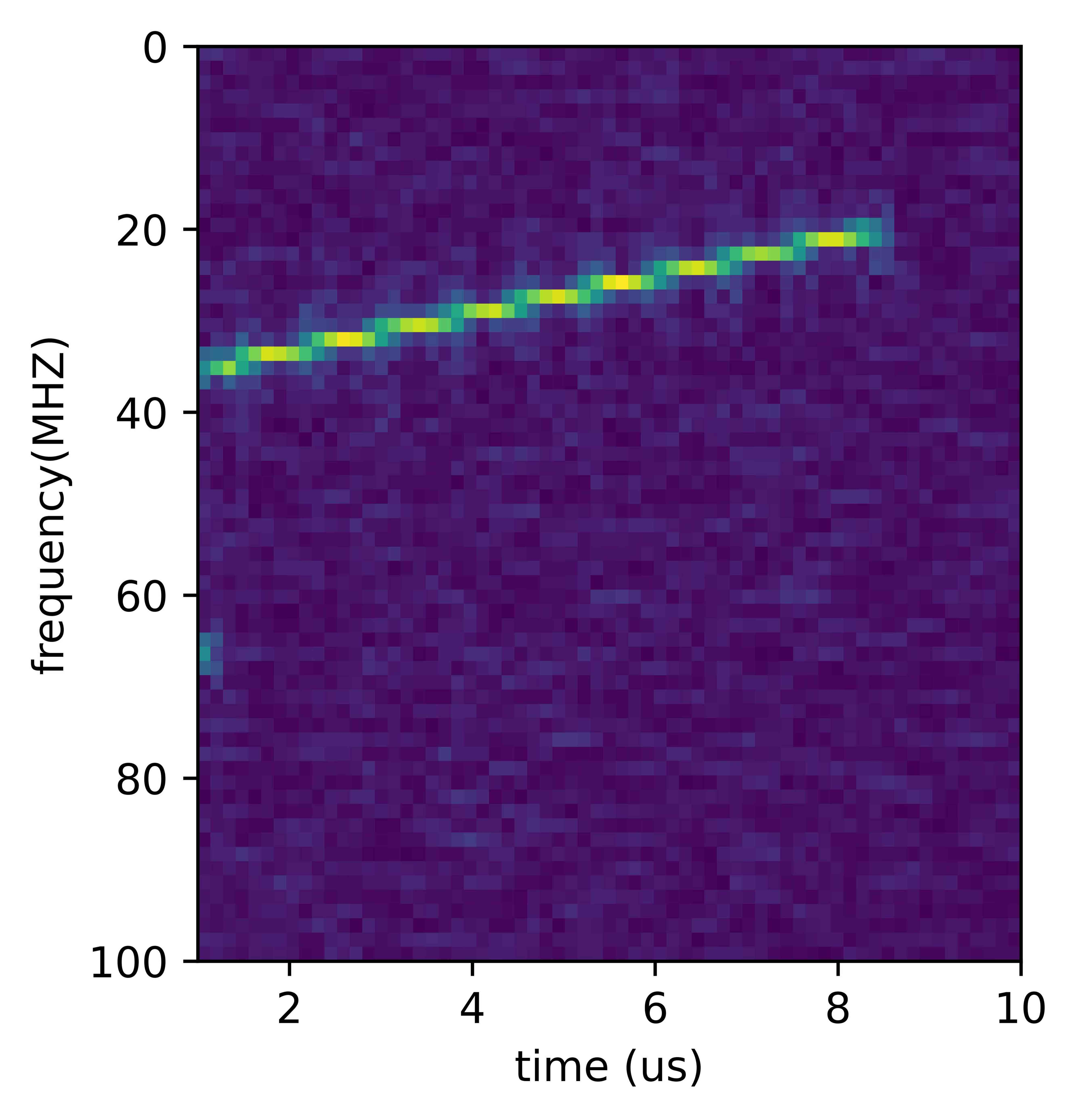}}
	\subfigure[Random frequency offset]{
		\includegraphics[width=0.3\textwidth]{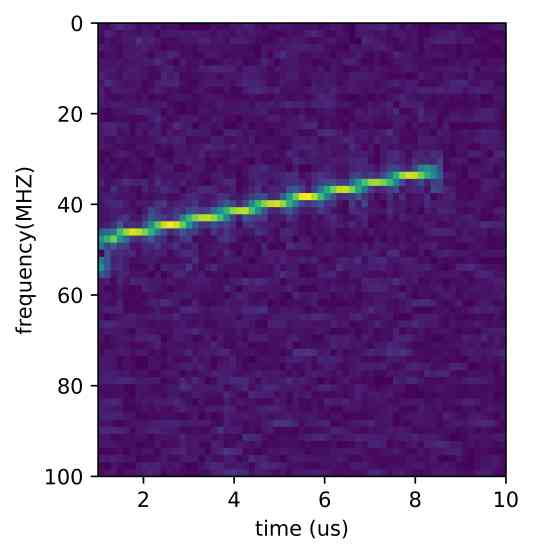}}
	\subfigure[Random additive white Gaussian noise]{
		\includegraphics[width=0.3\textwidth]{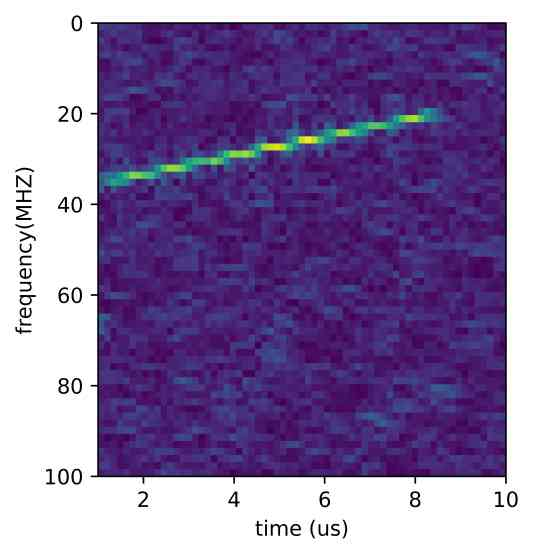}}
	\subfigure[Complex conjugate]{
		\includegraphics[width=0.3\textwidth]{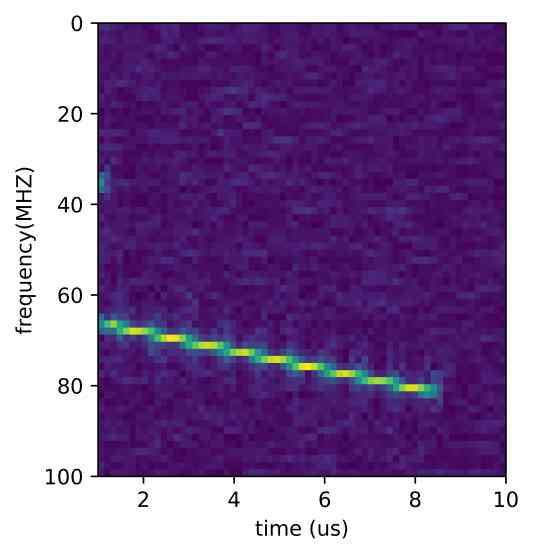}}		
	\subfigure[Random time masking]{
		\includegraphics[width=0.3\textwidth]{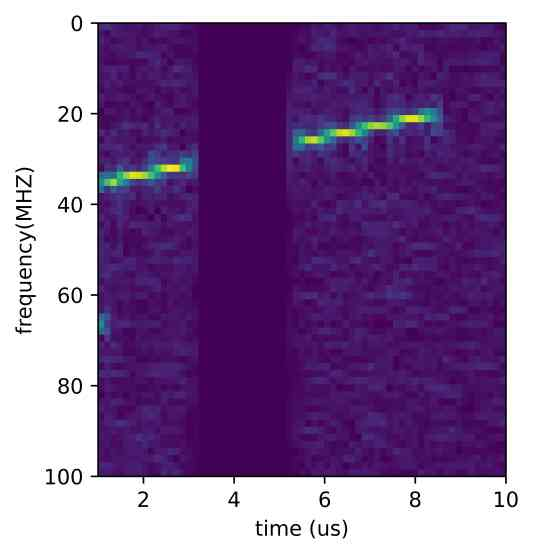}}
	\subfigure[Random resample]{
		\includegraphics[width=0.3\textwidth]{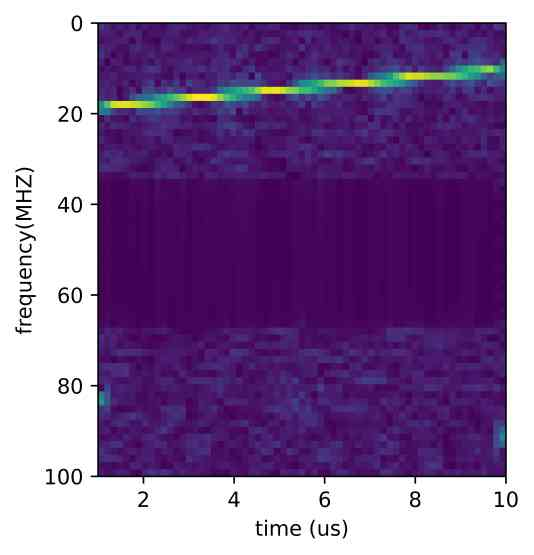}}
	\subfigure[Rayleigh-fading-channel]{
		\includegraphics[width=0.3\textwidth]{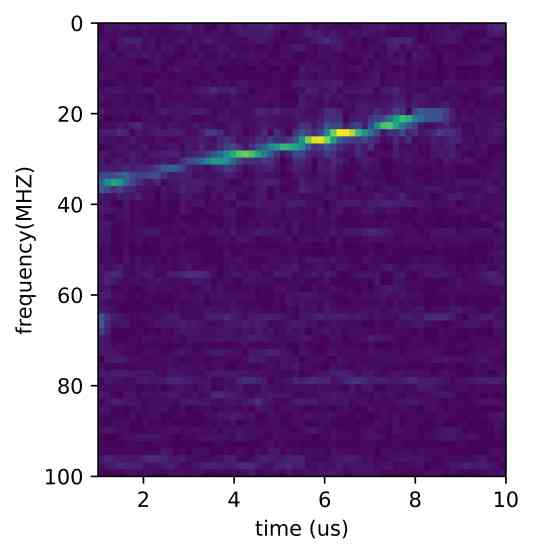}}	
	\caption{Effect of different augmentation methods on a LFM signal}
	\label{augmentation}
\end{figure}

In Eq.\ref{signal_model}, only phase function $\theta(k)$ determines the types of the waveforms, and therefore other parameters like 
$f_c$,$f_s$ can be viewed as class-independent parameters.Therefore,
 the following  augmentation methods are defined to alter those parameters to generate $\widetilde{y}(k)$ from $y(k)$.
\begin{enumerate}
	\item	\textbf{Random frequency offset\\} 
	\begin{align}
		\widetilde{y}(k) & = y(k)\cdot exp(j\cdot2\pi \frac{f_0}{f_s} k + \theta)
	\end{align}
	Where $f_0$, $\theta$ are random variables. This augmentation creates samples with different carrier frequency. In this way, radar signals with frequency agility are also simulated.\\
	\item	\textbf{Random additive white Gaussian noise\\}
	The augmented signal is 
	\begin{align}
		\widetilde{y}(k) & = y(k)+\gamma \cdot \widetilde{n}(k)
	\end{align}
	 $\gamma$ controls the intensity of the noise, $\widetilde{n}(k)$ is a  pseudo-random number array with Gaussian distribution.\\

	\item	\textbf{Complex conjugate\\}
	The augmented signal can be expressed as
	\begin{align}
		\widetilde{y}(k) & = Conj(y(k))
	\end{align}
	and $Conj(\cdot)$ represents a complex conjugate operator. This operation flips the phase function  $\theta(k)$ of the original signal because of the property of the Fourier transform  $FT(\widetilde{y}(k))(f)=Conj(FT(y(k))(-f)$.The such operation also only changes the class-independent parameters\\

	\item	\textbf{Random time masking\\} 
	Similar to cutout\cite{devriesImprovedRegularizationConvolutional2017a}
	that are used for image augmentation, a fragment of the signal is masked to be 0 as 
	\begin{align}
		\widetilde{y}(k) & = \left\{
			\begin{aligned}
				0 ,\quad   L < k < U\\
				y(k), \quad   othervise
			\end{aligned}
			\right.
	\end{align}
	where $L$ , $U$ is the start time, end time of the time mask respectively. This augmentation can alter the pulse width of $y(k)$, which is also recognized as a class-independent parameters \\

	\item	\textbf{Random resample\\}
	 Changing the sampling rate will not change the high-level features of the received signal as long as the changed sampling rate is greater than the Nyquist frequency. This process can be implemented as the cascading interpolation and decimation as
	 \begin{align}
		\widetilde{y}(m) & = decima(FIR(interpo(y(k),fs_h)),fs_L)
	\end{align}
	,where $fs_h$,$fs_L$ is the sampling rate after interpolation and decimation, $FIR(\cdot)$represents a low-pass filter to avoid aliasing.\\

	\item	\textbf{Rayleigh-fading-channel\\}
	The channel is also class-independent.To simulate the channel  effect on the signal, the sum-of-sinusoids statistical simulation model\cite{clarkeStatisticalTheoryMobileradio1968} is implemented as 
	\begin{align}
		h_{I}\left(k\right) = \frac{1}{\sqrt{M}} \sum_{m = 1}^{M} \cos \left\{2 \pi f_{D} \cos \left[\frac{(2 m-1) \pi+\theta}{4 M}\right] \cdot k+\alpha_{m}\right\} \\
		h_{Q}\left(k\right) = \frac{1}{\sqrt{M}} \sum_{m = 1}^{M} \sin \left\{2 \pi f_{D} \cos \left[\frac{(2 m-1) \pi+\theta}{4 M}\right] \cdot k+\beta_{m}\right\} \\
		h\left(k\right) = h_{I}\left(k\right)+j h_{Q}\left(k\right)
	\end{align}
	where $ \theta$, $\alpha_{m}$  and  $\beta_{m}$  are uniformly distributed over  $[0,2 \pi)$ and $f_{D}$ is the maximum Doppler spread.Therefore,the augmented signal can be calculated as
	\begin{align}
		\widetilde{y}(k) & = y(k) \otimes h(k)
	\end{align}

\end{enumerate}
Figure \ref{augmentation} shows an LFM signal and its augmented samples with different augmentation transforms. During network training, a pipeline of these transforms are randomly chosen as in \cite{limFastAutoAugment2019}.

\subsection{Pseudo-supervised contrastive training}
Naively applying K-means to the features extracted by the pretext model can lead to cluster degeneracy\cite{caronDeepClusteringUnsupervised2018}, which stops further improvement in clustering performance.Two major reasons may cause this problem.Firstly, in self-supervised contrastive training, a single positive sample is contrastive against the entire remainder of the batch. Where there are inevitable some positive samples. Therefore, samples of points belonging to the same class are also pushed apart in embedding space.Sencondly, the augmentation may not be able to alter all the class-independent parameters (such as the number of bits in phase coded waveform) and therefore, the pretext model may fail to filter out such features.

 To closely align samples from the same class, we propose to use the Supervised Contrastive loss\cite{khoslaSupervisedContrastiveLearning2020} with Pseudo labels. First, the K-means clustering is applied to the output of the pretext model to obtain $C$ clusters and their clustering centers. Next, based on the assumption that samples closing to each clustering center are likely to be the same class,  the top $K$ nearest data points around each clustering center are viewed as  'reliable' samples their labels are assigned according to the cluster numbers.
 
 With those reliable samples, the pretext model can be further fine-tuned with the supervised contrastive loss as in \ref{supervisedcontrastiveloss}.Here, $\tau$, 
 $\boldsymbol{e}_{i}$ has the same meaning as Eq \ref{self-supervisedcontrastiveloss}.$D_{pse}$ is the set of the reliable samples. $A(i) \equiv D_{pse} \backslash\{i\}$ represent the reliable samples except $i$  and $P(i) \equiv\left\{p \in A(i): \tilde{\boldsymbol{cluster}}_{p}=\tilde{\boldsymbol{cluster}}_{i}\right\}$ is the samples in $A(i)$ with the same cluster numbers as $i$.
 \begin{align}
	\label{supervisedcontrastiveloss}
	L_{supc} = -\sum_{i \in D_{pse}} \frac{1}{|P(i)|} \sum_{p \in P(i)} \log \frac{\exp \left(\boldsymbol{e}_{i} \cdot \boldsymbol{e}_{p} / \tau\right)}{\sum_{a \in A(i)} \exp \left(\boldsymbol{e}_{i} \cdot \boldsymbol{e}_{a} / \tau\right)}
\end{align}

\begin{figure}[htbp]
	\centering
	\includegraphics[width=0.8\textwidth]{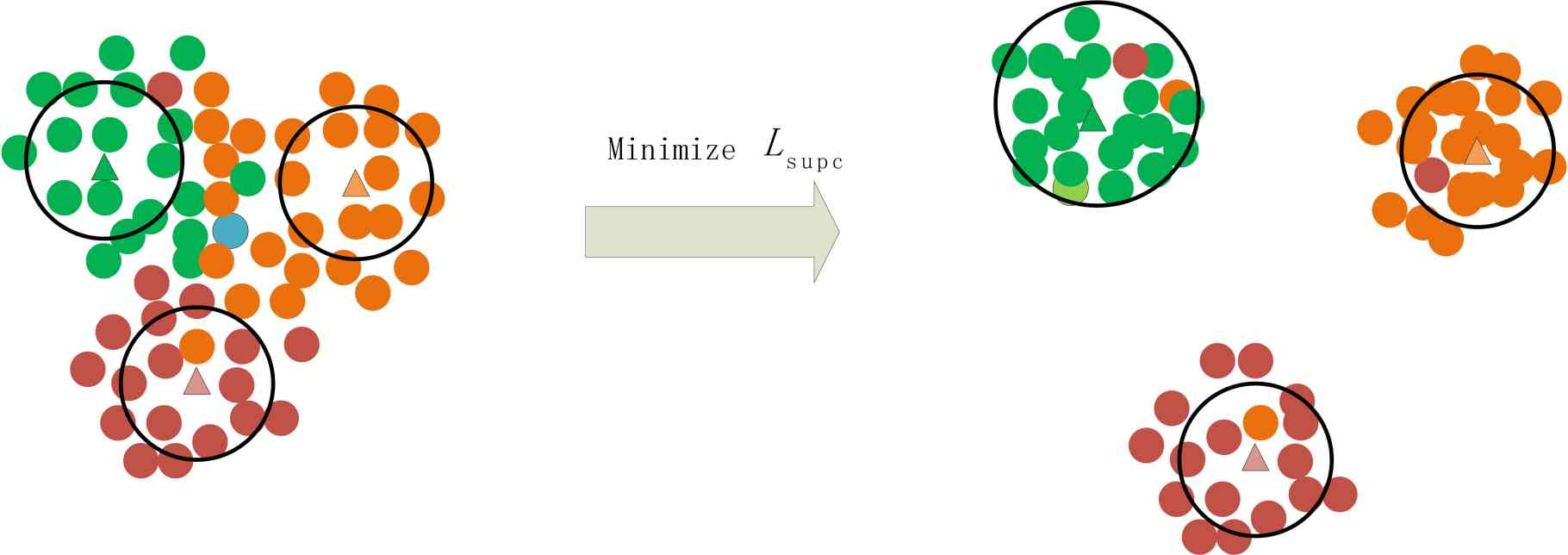}
	\caption{Effect of the Pseudo-supervised contrastive training, where different colors denote different clusters and the triangles denote the cluster centers produced by K-means. The points within each circle are considered reliable samples}
	\label{Pseudo-supervised}
\end{figure}

Figure \ref{Pseudo-supervised} illustrates the effect of Pseudo-supervised contrastive training on the embedding features produced by the pretext model. Through  this process, each cluster has been further separated while samples in the same cluster are pulled closer.     
 
\subsection{Semi-supervised self-labeling}
The Pseudo-supervised contrastive training has further separate samples in each cluster, however, it only uses the samples around the clustering centers of k-means while the majority of samples remain unlabeled. Inspired by the pseudo-label technique\cite{leePseudoLabelSimpleEfficient2013} in semi-supervised learning, we further convert the clustering problem into a semi-supervised problem.

\begin{figure}[htbp]
	\centering
	\includegraphics[width=0.8\textwidth]{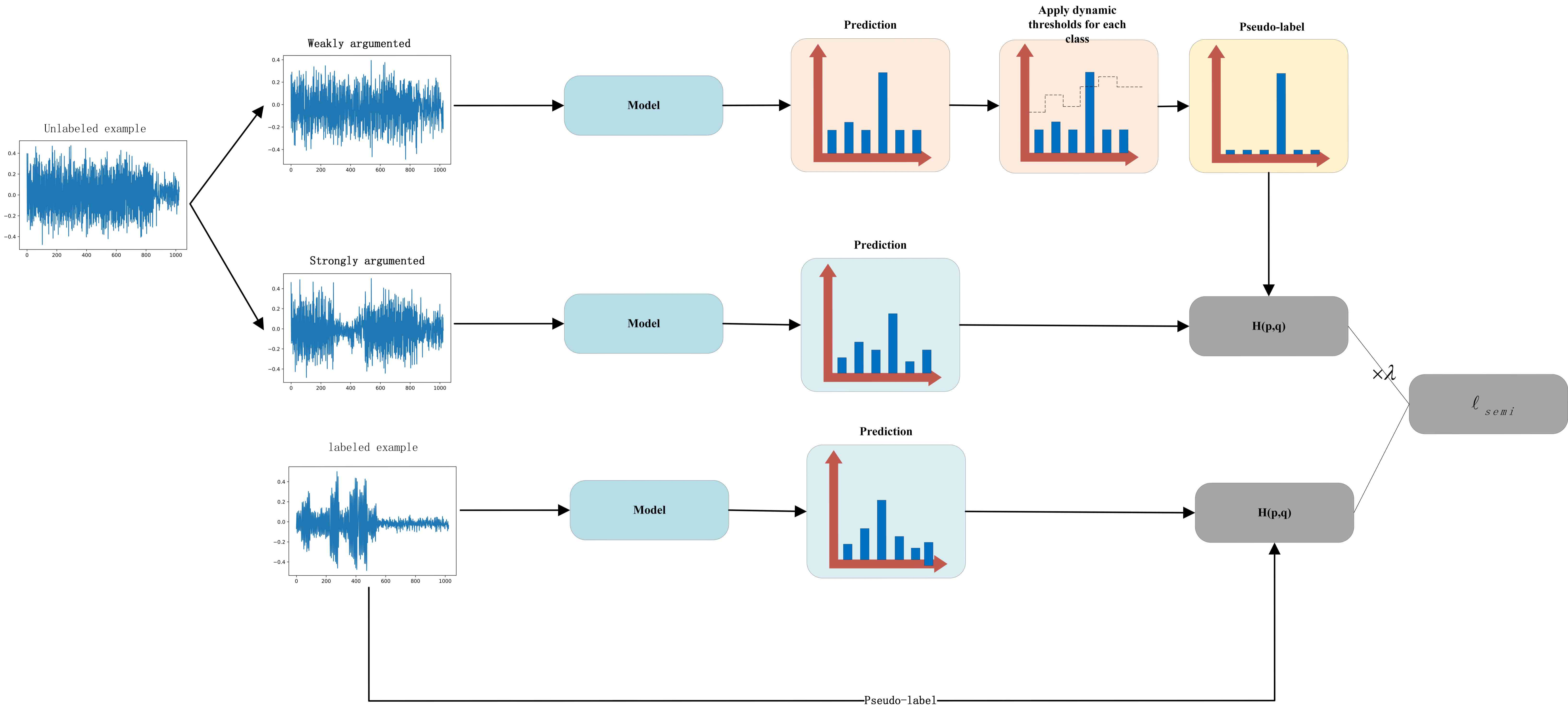}
	\caption{The forward path for calculating semi-supervised self-labeling loss}
	\label{semi-supervised}
\end{figure}

In particular, the K-means clustering algorithm is again performed on the fine-tuned pretext model to obtain the top $M$ nearest data points around each clustering center.Those samples are treated as labeled data points, and the rest samples remain unlabeled. Next, the model is trained to minimize the semi-supervised loss $\ell_{semi}$\cite{sohnFixMatchSimplifyingSemiSupervised2020}.
\begin{align}
	\ell_{s} = \frac{1}{N_{s}} \sum_{i = 1}^{N_{s}} \mathrm{H}\left(p_{i}, p_{\mathrm{m}}\left(y \mid x_{i}\right)\right)
	\label{superloss}
\end{align}

\begin{align}
	\ell_{u} & = \frac{1}{N_{us}} \sum_{i= 1}^{N_{us}} \mathbb{1}\left(\max \left(q_{i}\right) \geq \tau_{argmax(q_{i})}\right) \mathrm{H}\left(\hat{q}_{i}, p_{\mathrm{m}}\left(y \mid \mathcal{A}\left(x_{i}\right)\right)\right)
	\label{unsuperloss}
\end{align}
and
\begin{align}
	\ell_{semi} & = \ell_{s} + \lambda \ell_{u}
\end{align}
Here $\ell_{s}$ ,$\ell_{u}$ represent the supervised, unsupervised loss,$\lambda$ is the weight for the unsupervised loss.In Eq.\ref{superloss}, $p_{\mathrm{m}}\left(y \mid x_{i}\right)=softmax({{{\Phi }_\theta }\left( {{x_i}} \right)})$,represent the estimated probability distribution for the reliable sample ${x_i}$,$H(.)$ represents the standard cross- entropy loss, $p_{i}$ is the pseudo-label for $x_{i}$, $N_{s}$ is the batch size for labeled samples.In Eq.\ref{unsuperloss}, $q_{i}$ is the model estimated probability distribution for the $ith$ weakly augmented sample,$\mathcal{A}$ represents the strong data augmentation , $N_{s}$ is the batch size for the labeled samples and $\tau_{argmax(q_{i})}$
is the threshold for selecting confident samples .The process can be summarized in Figure.\ref{semi-supervised}.

The choice of the threshold $\tau_{argmax(q_{i})}$ is crucial for semi-supervised learning. Larger threshold could filter out noisy samples,however it may prevent learning for hard-to-learn classes.Therefore, the FlexMatch\cite{zhangFlexMatchBoostingSemiSupervised2021} algorithm is adopted to dynamically adjust the threshold during training.

\subsection{The design of the encoder}
\begin{figure}[htbp]
	\centering
	\includegraphics[width=0.8\textwidth]{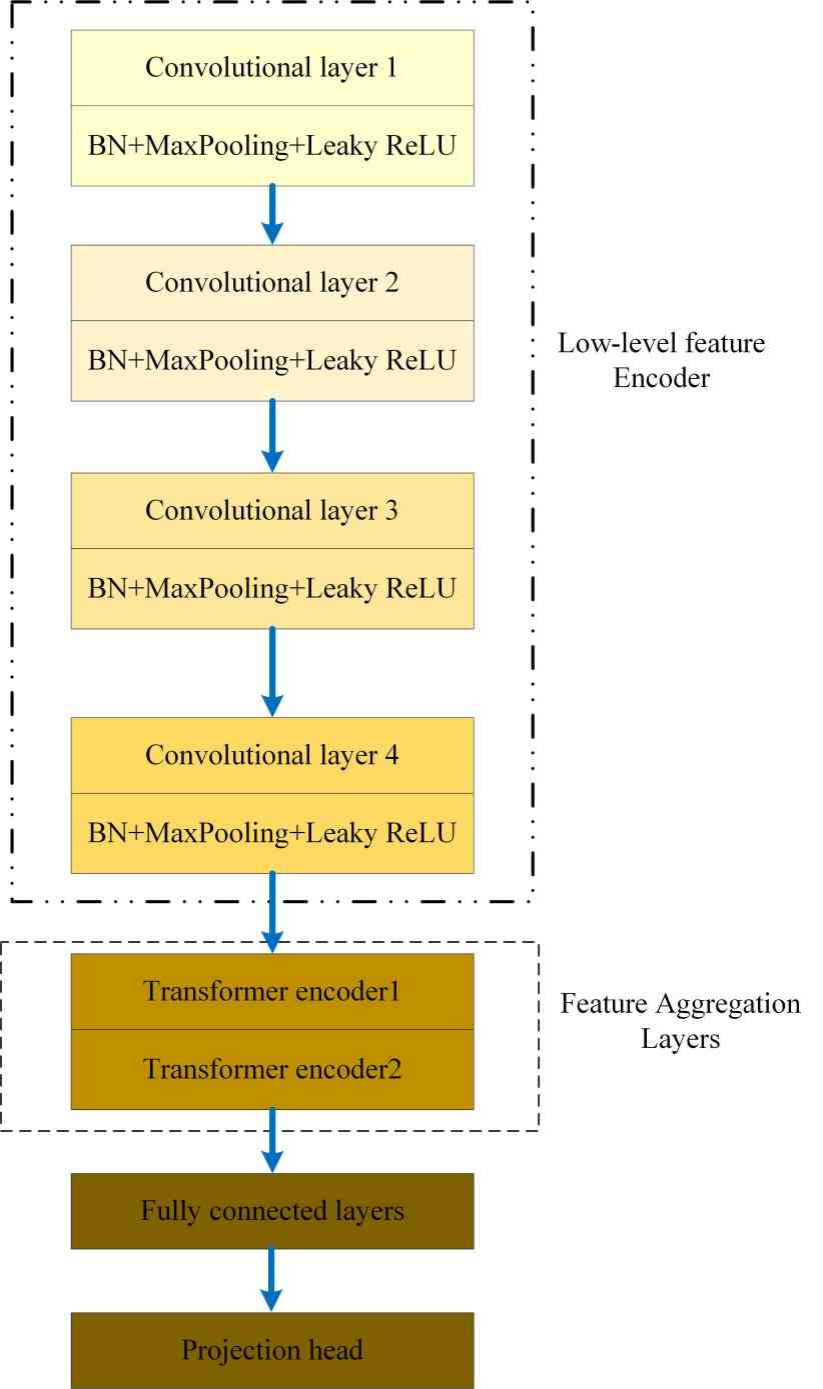}
	\caption{The architecture of the proposed encoder}
	\label{network_architecture of Trans-CNN}
\end{figure}

So far, we have discussed the training method for radar signal clustering, another factor that can heavily affect the clustering performance is the structure of the deep network. Many of the previous works \cite{wangTransferredDeepLearning2019,westDeepArchitecturesModulation2017,liuTERASelfSupervisedLearning2021}have suggested adding a layer to model the temporal relations between features could improve the classification performance , therefore, we design our encoder network(called Trans-CNN) by adding 2 Transformer Encoders\cite{vaswaniAttentionAllYou2017} layers over standard convolution blocks, as shown in \ref{network_architecture of Trans-CNN}.

Here, the stack of convolution blocks work as a group of digital filters to extract low-level feature from the raw IQ inputs and the Transformer Encoders uses the self-attention mechanism to extract contextual information.Between each convolution block, pooling layers are added to reduce the computational cost and the feature maps are further normalized by the Batch Normalization\cite{ioffeBatchNormalizationAccelerating2015} layers to speed up convergence.
The projection head is simply a linear layer and during inference, it is discarded as suggested by \cite{chenSimpleFrameworkContrastive2020}. 


\section{Results}
\label{Results}
In this section, several datasets are simulated  to test the proposed method and other baseline methods for different  metrics under various conditions.All the experiments were conducted on a laptop equipped with an AMD 4800h CPU, and an Nvidia RTX 2060 GPU.
The data augmentation and network optimization were conducted on Pytorch\cite{paszkePyTorchImperativeStyle2019}. 
\subsection{Dataset and training  settings}
Typical radars operate at the frequency ranging from 0.35 GHz to 18Ghz, and the receivers typically adopt a channelized structure to cover such wide bandwidth.The high-speed Analog-to-Digital Converter(ADC) of the receiver often operates at several GHz, but after digital channelization, the carrier frequency and the sampling rate has been reduced to tens to hundreds MHZ.Therefore, we assume the received radar signals are sampled with an equivalent sampling rate of 100MHZ. 12 typical radar signals are generated namely linear frequency modulation(LFM),none-linear frequency modulation(NLFM),barker phase code (Barker),three frequency codes (CostasFM,2-FSK,4-FSK),four polytime codes (T1, T2, T3, and T4) as well as two composite modulations(LFM-BFSK, LFM-p1 code).The detailed parameters for each type are summarized in Table \ref{parameters_dataset}.We assume about half of the samples are received through Rayleigh fading channels(in Eq.\ref{channel}, $p=0.5$). The channel conditions are set according to \cite{huynh-theAccurateLPIRadar2021}.
\begin{table}[]
	\resizebox{\textwidth}{!}{
	\begin{tabular}{llll}
	\toprule
	\textbf{Type} &
	\textbf{Carrier Frequency (MHZ)} &
	\textbf{Pulse width (us)} &
	\textbf{Parameters} \\
	\hline
	LFM &
	  U(10,20) &
	  U(5,10) &
	  \begin{tabular}[c]{@{}l@{}}Bandwidth \\ U(10,20)   MHZ, SweepDirection, random choose from ‘up’ or ‘down’.\end{tabular} \\
	\hline
	NLFM &
	  U(10,20) &
	  U(5,10) &
	  \begin{tabular}[c]{@{}l@{}}Bandwidth\\  U(10,20)   MHZ\end{tabular} \\
	\hline
	BPSK &
	  U(10,20) &
	  U(5,10) &
	  7,11,13-bit   Barker code \\
	\hline
	CostasFM &
	  U(10,20) &
	  U(5,10) &
	  Number of Frequency hop {[}5 7 10{]} \\
	\hline
	2FSK &
	  U(10,20) &
	  U(5,10) &
	  Frequency separation   U(5, 10) MHZ \\
	\hline
	4FSK &
	  U(10,20) &
	  U(5,10) &
	  Frequency separation   U(5, 10) MHZ \\
	\hline
	T1, T2 &
	  U(10,20) &
	  U(5,10) &
	  \begin{tabular}[c]{@{}l@{}}Number of   phase state: 2 \\ Number of   segments:\\   {[}18,20,22{]}\end{tabular} \\
	\hline
	T3 ,T4 &
	  U(10,30) &
	  U(5,10) &
	  \begin{tabular}[c]{@{}l@{}}Number of   phase state: 2\\     Number of   segments:\\   {[}18,20,22{]}\\   Bandwidth\\  U(5,10)  MHZ\end{tabular} \\
	\hline
	LFM-2FSK &
	  U(10,30) &
	  U(5,10) &
	  \begin{tabular}[c]{@{}l@{}}Bandwidth\\    U(10,20)   MHZ,\\   Frequency separation   U(5, 10) MHZ\end{tabular} \\
	\hline
	LFM-bpsk &
	  U(10,20) &
	  U(5,10) &
	  \begin{tabular}[c]{@{}l@{}}Bandwidth\\  U(10,20)   MHZ,\\   4,9,16-bit Barker code
	\end{tabular}\\
	\bottomrule
	\end{tabular}
	}
	\caption{Parameters for each waveform }
	\label{parameters_dataset}
\end{table}
Two datasets are generated according to Table.\ref{parameters_dataset},the first one contains 1000 signals per waveform with the SNR ranging from 5dB to 15dB to train and test the performance of the proposed model along with other baseline methods.The SNR of the second dataset varies from -10 dB to +10 dB with a step of 1.0 dB , and for each SNR level, 200 samples are generated for each waveform type and it is used for testing only. So there are in total 12000 samples for dataset 1 and 50400 samples for dataset 2.

The parameters for each type of data augmentation is summarized in Table\ref{parameters_augmentation}.A probability is introduced to control the intensity of the augmentation.As shown in the table, we adopt the strongest augmentation for contrastive training,and a moderate-intensity augmentation, weak augmentation for the semi-supervised self-labeling.

\begin{table}[]
	\resizebox{\textwidth}{!}{
	\begin{tabular}{lllll}
	\toprule
	Types of augmentation &
	  Parameters &
	  \begin{tabular}[c]{@{}l@{}}Selection \\ probability \\ (weak aug)\end{tabular} &
	  \begin{tabular}[c]{@{}l@{}}Selection \\ probability \\ (moderate aug)\end{tabular} &
	  \begin{tabular}[c]{@{}l@{}}Selection \\ probability \\ (strong aug)\end{tabular} \\ \hline
	Random frequency offset & F0                                       & 0.5 & 0.5 & 0.5 \\ \hline
	\begin{tabular}[c]{@{}l@{}}Random additive white \\ Gaussian noise\end{tabular} &
	  \begin{tabular}[c]{@{}l@{}}noise power :  0.3679 \textasciitilde 1.6487 \\ times the signal power\end{tabular} &
	  0.5 &
	  1 &
	  1 \\ \hline
	Complex conjugate       & None                                     & 0.5 & 0.5 & 0.5 \\ \hline
	Random time masking     & Masking size : 100 \textasciitilde 300 data points       & 0.5 & 0.5 & 0.5 \\ \hline
	Random resample         & Resample frequency range : 50MHZ \textasciitilde 150MHZ & 0   & 0.3 & 0.5 \\ \hline
	Rayleigh-fading-channel & number of sinusoids : 32                 & 0   & 0.3 & 0.5 \\ \bottomrule
	\end{tabular}
	}
	\caption{Parameters for data augmentation}
	\label{parameters_augmentation}
\end{table}

Table \ref{Hyperparameter_network} summerizes some important hyperparameters of the proposed network. The hyperparameters of the convolutional layers are set according to  \cite{jingAdaptiveFocalLoss2022}.For all three training stages,the batch sizes are set to  512, and some important configurations for each training stage are listed in table\ref{training_configuration}.The training stopped when the loss no longer decrease.

\begin{table}[]
	\begin{tabular}{ll}
	\toprule
	\textbf{Layers}         & \textbf{Parameters}                                         \\ \hline
	Conv1                   & Kernel Size 1 × 5, channel 32                               \\ \hline
	Conv2                   & Kernel Size 1 × 5, channel 64                               \\ \hline
	Conv3                   & Kernel Size 1 × 5, channel 128                              \\ \hline
	Conv4                   & Kernel Size 1 × 3, channel 256                              \\ \hline
	Transformer   encoder 1 & number of heads =8, dimension of the feedforward network=128 \\ \hline
	Transformer   encoder 2 & number of heads =8, dimension of the feedforward network=128 \\ \hline
	Fully connected  layer  & output dimension= 128                                       \\ \hline
	Projection head         & output dimension= 12                                        \\
	\bottomrule
\end{tabular}
\caption{Hyperparameters for proposed network}
\label{Hyperparameter_network}
\end{table}

\begin{table}[]
	\resizebox{\textwidth}{!}{
	\begin{tabular}{llll}
	\toprule
	\textbf{Training stage name}  & \textbf{optimizer} & \textbf{Learning   rate} & \textbf{Other param} \\ \hline
	Contrastive training & SGD     & 1e-4     & Temperature=1 \\ \hline
	Pseudo-supervised   contrastive training & SGD  & 1e-4 & \begin{tabular}[c]{@{}l@{}}Temperature=1\\ Number of nearest neighbors=200\end{tabular}           \\ \hline
	Semi-supervised   self-labeling          & ADAM & 2e-4 & \begin{tabular}[c]{@{}l@{}}Maximum threshold =0.99\\  Number of nearest neighbors=300  \\ $\lambda=1$ \end{tabular} \\
	\bottomrule
\end{tabular}
	}
\caption{Training configuration for each stage}
\label{training_configuration}
\end{table}

\subsection{The quantitive  result on the first dataset}
The performance of the proposed method was tested and compared with some state-of-the-art deep clustering algorithm including SimCLR\cite{chenSimpleFrameworkContrastive2020} SCAN\cite{vangansbekeSCANLearningClassify2020}, autoencoder(AE)\cite{xieUnsupervisedDeepEmbedding2016}, N2D\cite{mcconvilleN2DNotToo2021}with UMAP\cite{mcinnesUMAPUniformManifold2020},InfoGan\cite{chenInfoGANInterpretableRepresentation2016,gongUnsupervisedSpecificEmitter2020}. The supervised learning methods(with or without data augmentation ) using the network proposed in \cite{jingAdaptiveFocalLoss2022} was also added as the performance upper bound.Here only the weak augmentation in Table \ref{parameters_augmentation} was adopted as we found the strong and moderate augmentation got even worse performance, which is consistent with the founding in \cite{chenSimpleFrameworkContrastive2020}.For the models that output features instead of classes(like SimCLR), follow most of the literature, the k-means clustering is performed on the output features.The results are evaluated based on clustering accuracy (ACC), normalized mutual information (NMI), and adjusted rand index (ARI).
The scores for each metric are summarized in table \ref{results_1}
\begin{table}[]
	\resizebox{\textwidth}{!}{
	\begin{tabular}{llllllllll}
	\toprule
	 &
	  \textbf{DRSC} &
	  \textbf{SCAN} &
	  \textbf{SimCLR(CNN)} &
	  \textbf{Supervised} &
	  \textbf{Supervised (data aug)} &
	  \textbf{SimCLR} &
	  \textbf{Ae} &
	  \textbf{N2d} &
	  \textbf{Infogan} \\ \hline
	\textbf{Acc} & 97.05±0.77 & 92.03±2.18 & 79.40±0.87 & 97.32±0.42 & 97.81±0.34 & 95.09±0.01 & 20.12±1.24 & 21.34±0.65 & 26.07±5.90 \\ \hline
	\textbf{Ari} & 94.51±1.13 & 89.83±3.25 & 72.13±0.65 & 94.36±0.86 & 95.31±0.72 & 91.25±0.02 & 7.34±0.66  & 7.34±0.66  & 13.68±5.19 \\ \hline
	\textbf{Nmi} & 96.52±0.53 & 94.55±1.47 & 81.57±0.31 & 95.32±0.81 & 95.78±0.45 & 94.63±0.01 & 20.31±1.24 & 20.31±1.24 & 30.19±7.15 \\ 
	\bottomrule
	\end{tabular}
	}
	\caption{Clustering metrics (\%) for different methods . For each method, the experiment was repeated 6 times to calculate the average and standard deviation.}
	\label{results_1}
\end{table}

The results show that directly applying k-means on the proposed pretext model trained with the proposed data augmentation methods can already outperforms AE and GAN based method by a large margin.With another 2-stage of fine-tuning, the proposed DRSC get around $2\%$ improvement on all three metrics, which is very close to the supervised method without data augmentation. Noticeable, in contradiction to the result on image clustering, SCAN even perform worse than the  pretext model and the reason can be found in section \ref{analysis_features}.Using the weak augmentation, yields a 0.5 \textasciitilde 1 $\%$ performance improvement. Another important finding is that the backbone structure heavily affects the performance of the contrastive training as the CNN performs 15 \textasciitilde 20 $\%$ worse than the proposed Trans-CNN.

\subsection{Performance under different SNR}
In this section, we pick out 4 best-performing methods trained on the first dataset and test using the second dataset.The ACC,NMI,ARI were calculated for every SNR.
\begin{figure}[htbp]
	\centering
	\includegraphics[width=0.8\textwidth]{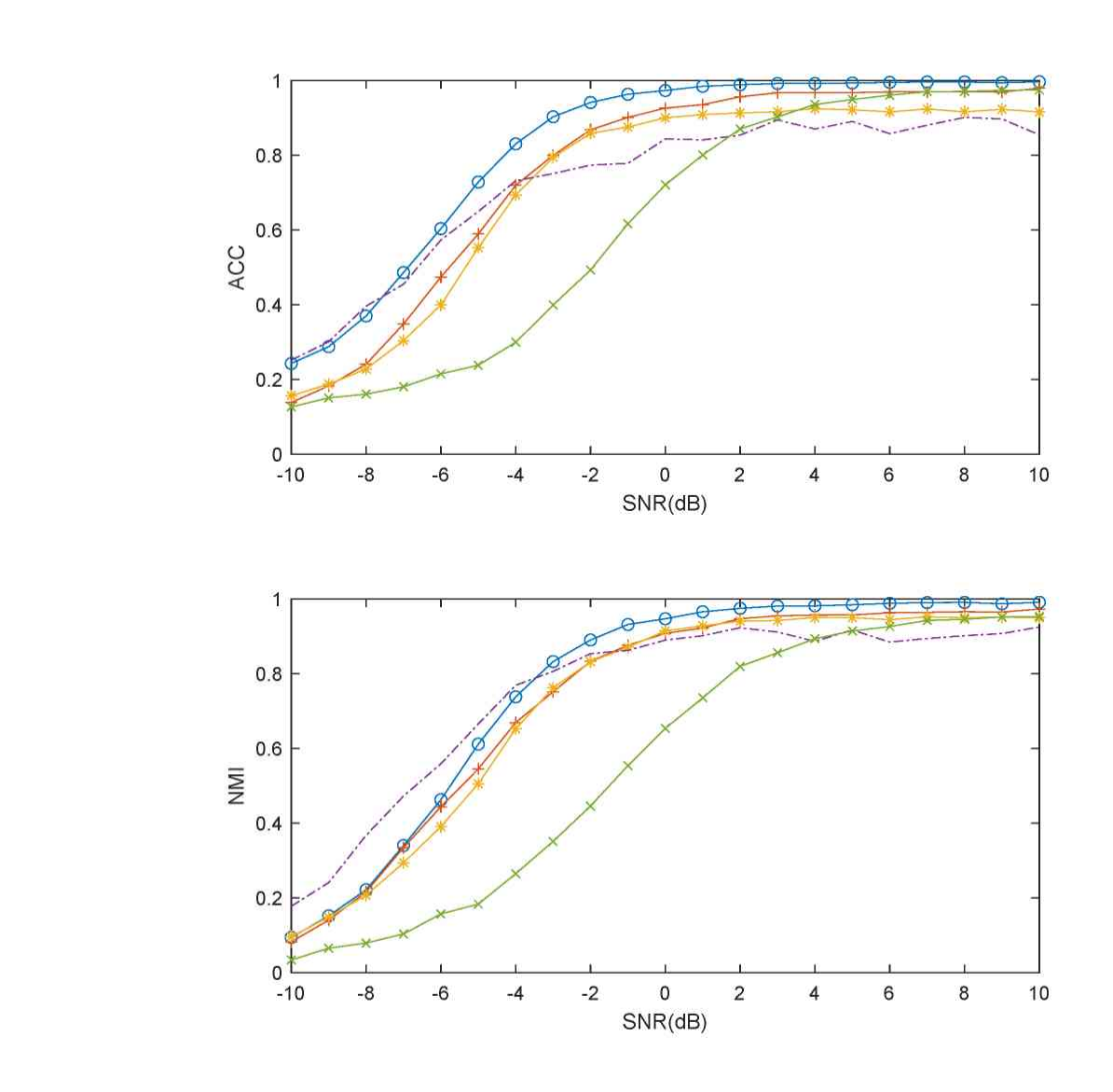}
	\caption{The result of the clustering metrics under different SNR}
	\label{snr_result}
\end{figure}
Figure\ref{snr_result} shows 3 clustering metric for each method, the proposed DRSC achieve an overall best performance among all the unsupervised methods, even outperfroms the supervised training without data augmentation under low SNR. It is also observed that the data augmentation significantly increases the model's robustness against noise as it can bring up to 50 $\%$ of performance boost for the supervised method when SNR $\le$ 0 dB.

\section{Discussion}
\label{Discussion}

In this section, we would like to further analyze the output features of each model and evaluate the model's performance under general conditions where the number of clusters is unknown.

\subsection{analysis of the output features}
\label{analysis_features}
\begin{figure}[htbp]
	\centering
	\subfigure[original signal]{
		\includegraphics[width=0.3\textwidth]{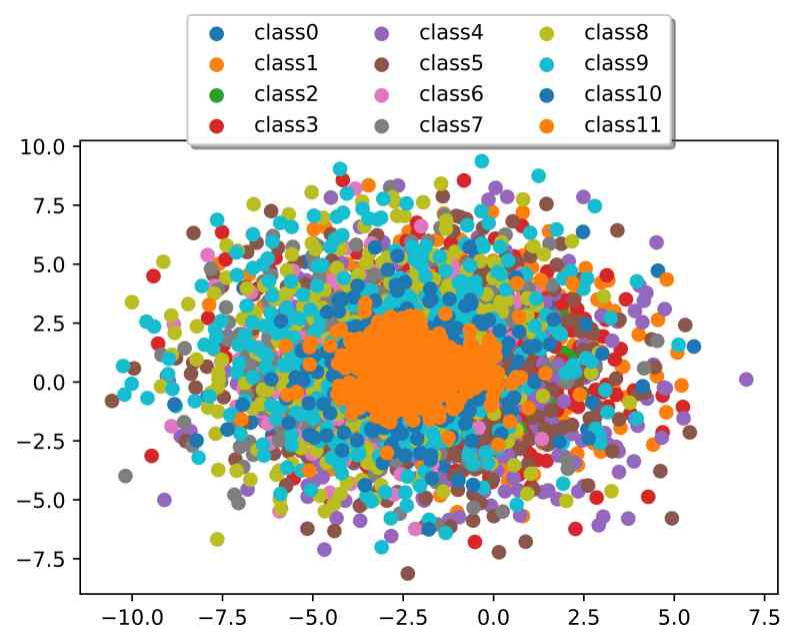}}
	\subfigure[Autoencoder]{
		\includegraphics[width=0.3\textwidth]{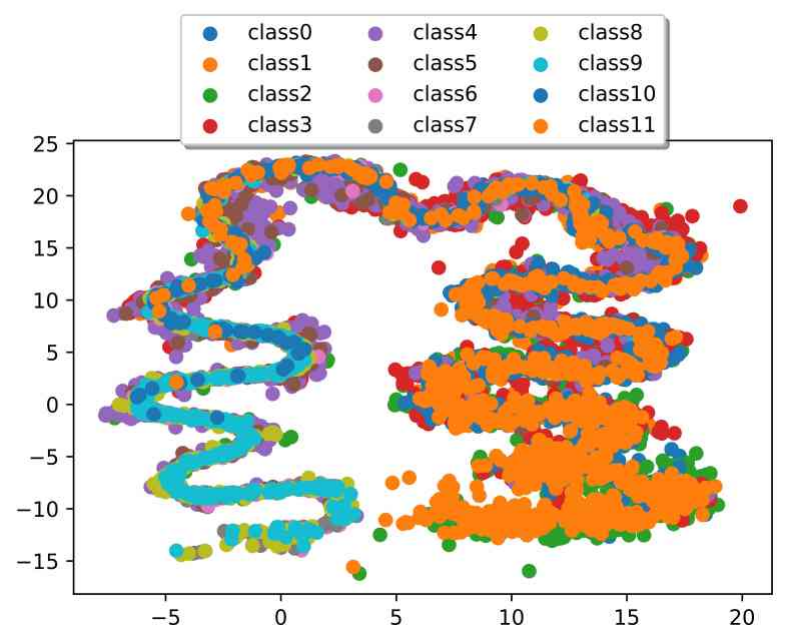}}
	\subfigure[SimCLR]{
		\includegraphics[width=0.3\textwidth]{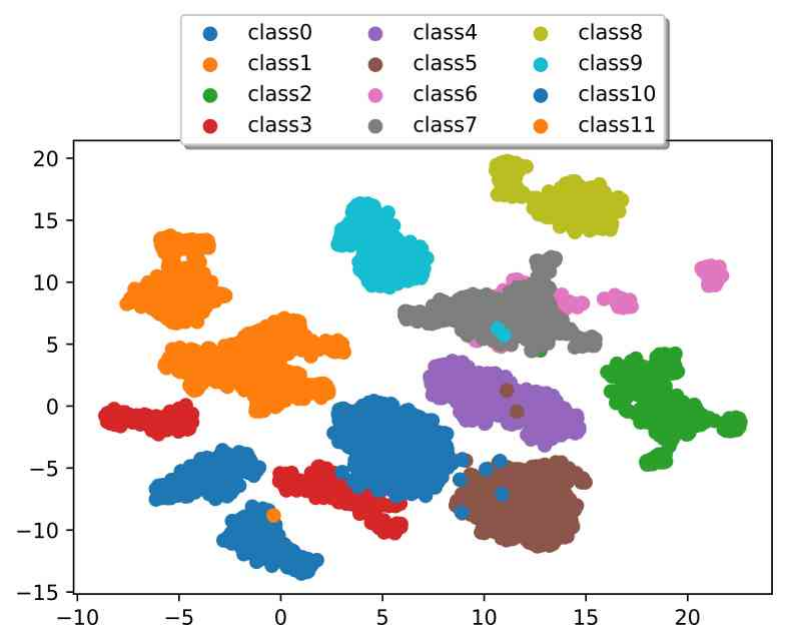}}
	\subfigure[SCAN]{
		\includegraphics[width=0.3\textwidth]{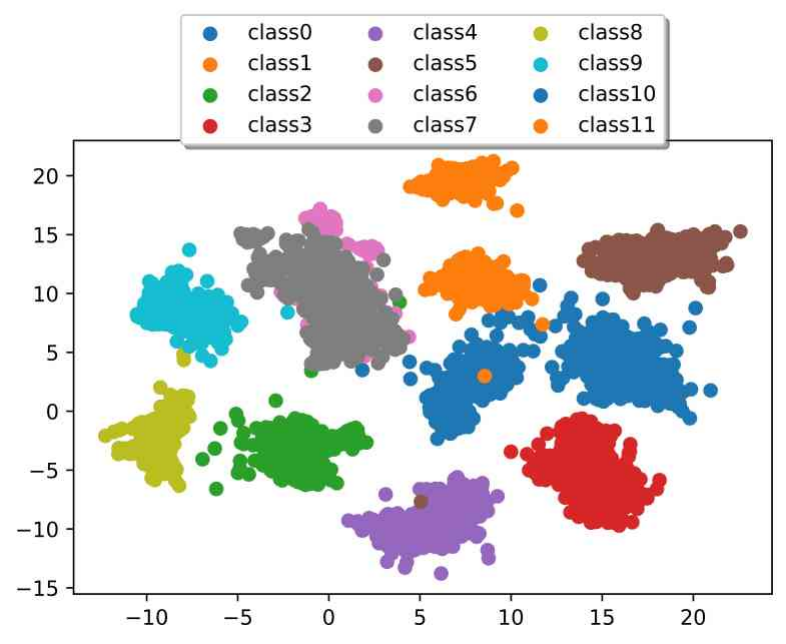}}		
	\subfigure[DRSC]{
		\includegraphics[width=0.3\textwidth]{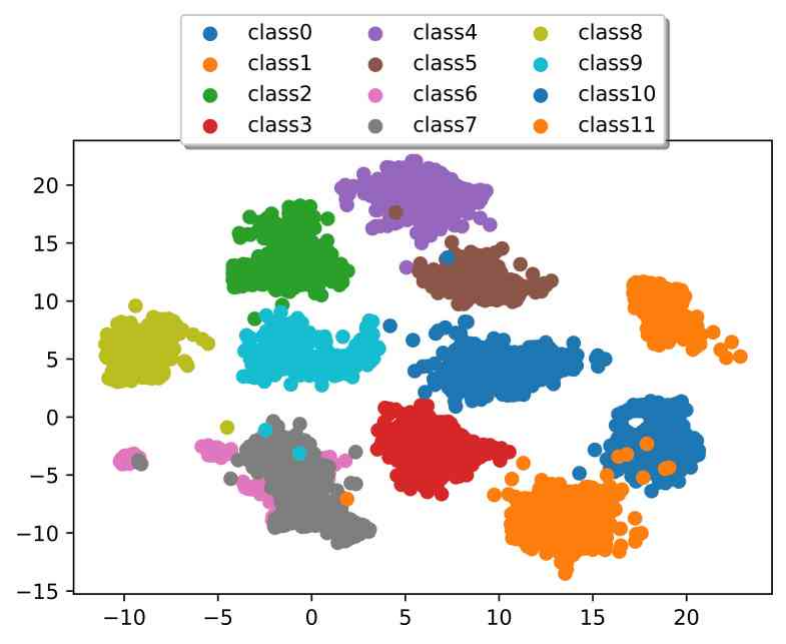}}
	\subfigure[Supervised(with data aug)]{
		\includegraphics[width=0.3\textwidth]{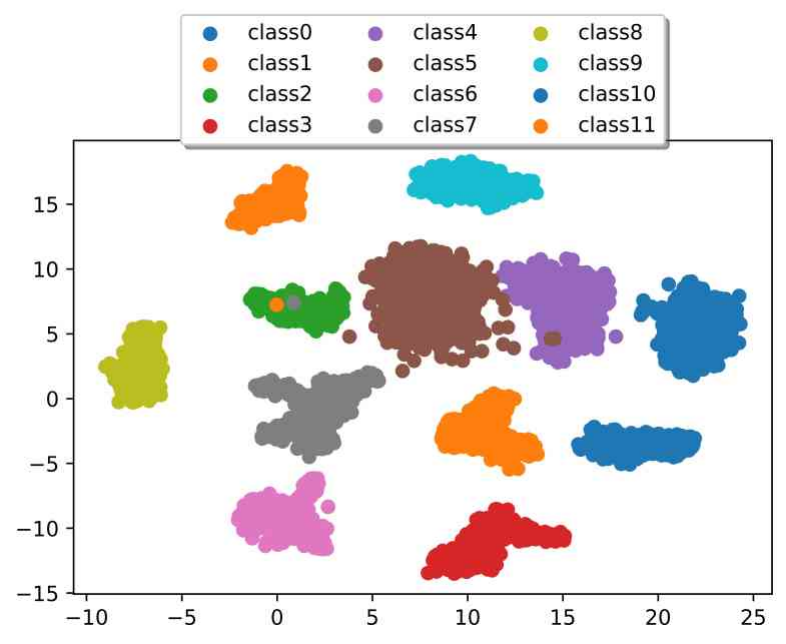}}	
	\caption{The 2D feature maps produced by UMAP on dataset 1.Class 0 \textasciitilde 11 represent LFM,NLFM,BPSK,CostasFM,2FSK,4FSK,T1,T2,T3,T4,LFM-2FSK,LFM-BPSK respectively.}
	\label{visualize_features}
\end{figure}

\begin{figure}[htbp]
	\centering
	\subfigure[Supervised]{
		\includegraphics[width=0.2\textwidth]{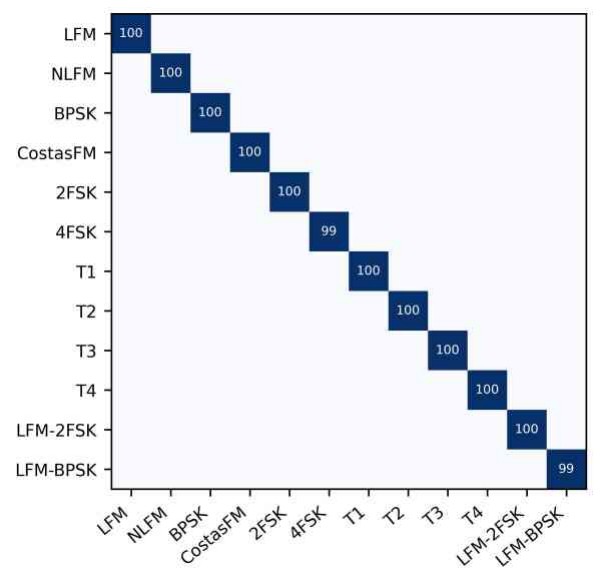}}
	\subfigure[SCAN]{
		\includegraphics[width=0.2\textwidth]{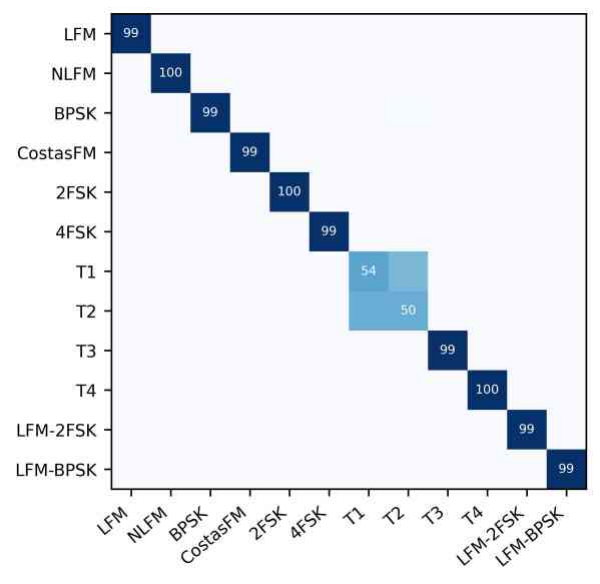}}
	\subfigure[SimCLR]{
		\includegraphics[width=0.2\textwidth]{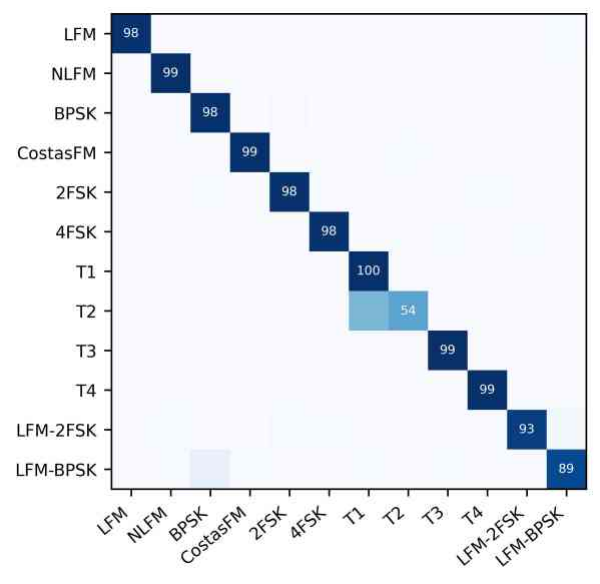}}
	\subfigure[DRSC]{
		\includegraphics[width=0.2\textwidth]{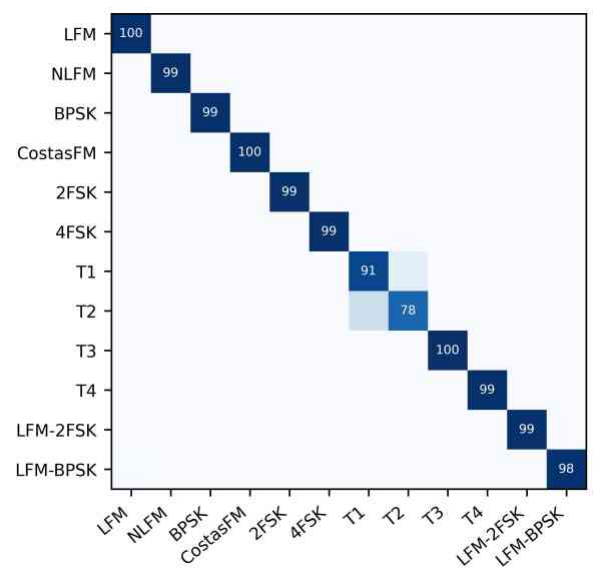}}		
	\caption{The confusion matrixes for different methods on dataset 1}
	\label{confusion_matrix}
\end{figure}

To understand the effectiveness of the proposed method on feature extraction, the dimensionalities of the output features of each model are reduced and visualized using the UMAP on the first dataset. 
As in Figure \ref{visualize_features}, 
Figure\ref{visualize_features} provides 2D-UMAP features for each case. Without any supervision,the contrastively trained model can effectively extract high-level class-dependent features to form distinctive clusters. The proposed method have larger degree of between class discrimination and within class aggregation compared with the pretext model, SimCLR, and SCAN,proving the effectiveness of pseudo-supervised contrastive training.Noticeable, the supervised method still has the advantage of splitting signals with similar features( the T1 T2), which can be seen from the confusion plots(Figure \ref{confusion_matrix}).

\begin{figure}[htbp]
	\centering
	\includegraphics[width=0.4\textwidth]{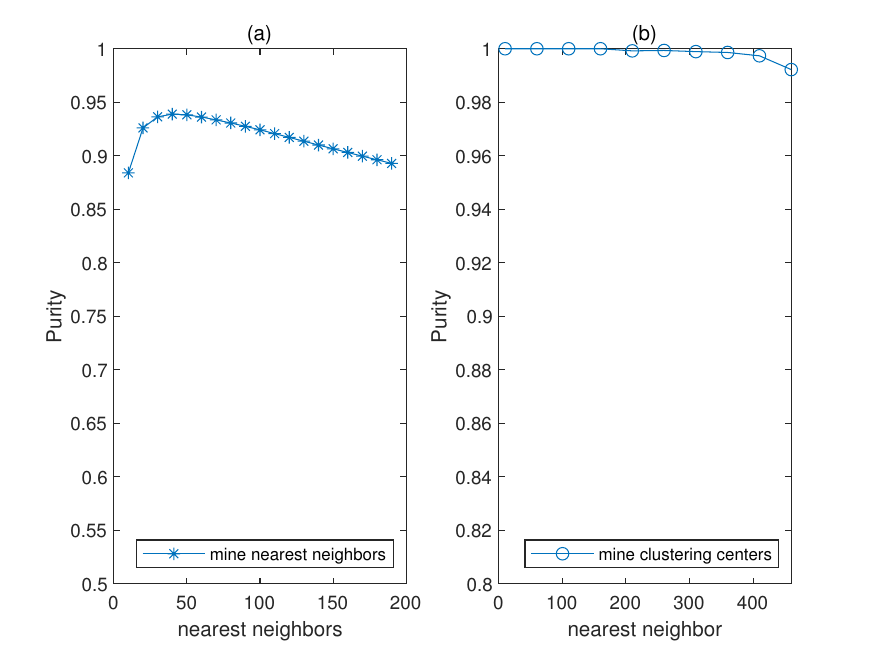}
	\caption{The result of two strategies of obtaining the pseudo labels with different number of nearest neighbors (a)mining the nearest neighbors (b) mining the clustering centers}
	\label{nearest_neighbors}
\end{figure}

In the previous sections,the simulations reveal that the performance of SCAN is  worse than SimCLR, which is contradictory to the founding in the original paper and the reason can be found by analyzing the pseudo labels produced by the pretext model.
Figure \ref{nearest_neighbors} quantifies the average probabilities that the nearest neighbors are instances of the same radar signal classes.

We could conclude that the performance degeneration of SCAN is due to a relatively low purity of the nearest neighbors, the highest purity is even lower than 95 $\%$.On the contrary, mining the clustering centers produces nearly perfect pseudo labels.

\subsection{Situation of an unknown number of clusters}
\begin{figure}[htbp]
	\centering
	\includegraphics[width=0.4\textwidth]{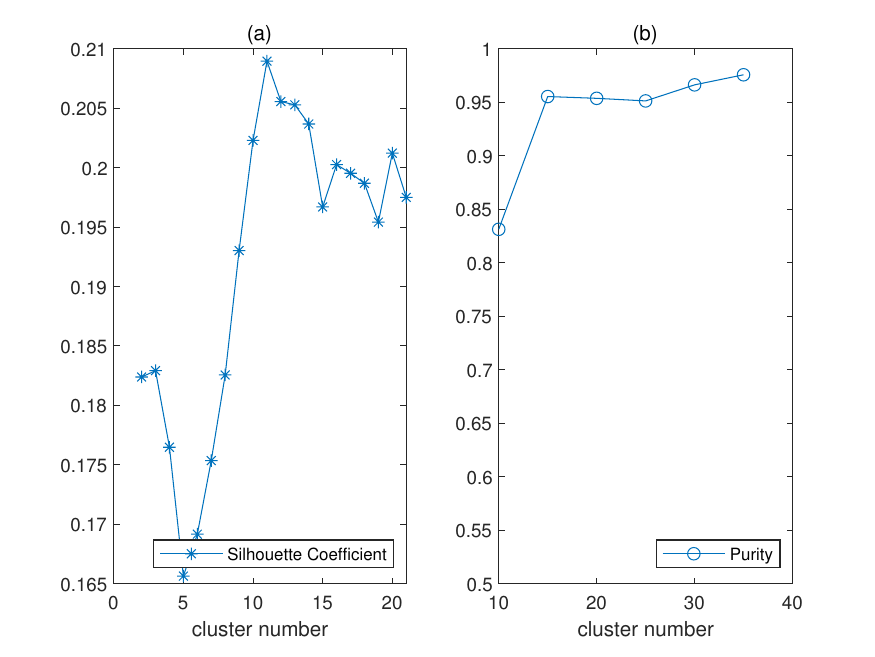}
	\caption{The Silhouette coefficient of the pretext model\textbf{(a)} and the purity of the proposed method \textbf{(b)} under different number of clusters}
	\label{overclustering_result}
\end{figure}

In the above discussions, the number of ground-truth classes is assumed to be known.However, in a real-world situation, the exact number of classes in the dataset may be  unknown.To further explore the limit of the proposed method, the experiments were re-conducted for the number of clusters and Figure \ref{overclustering_result} reports the results.If the arguments of the maxima of the Silhouette coefficient is used to estimate the number of classes, the estimated classe number is 11, which is just one less than the ground truth. Interestingly, as the number of clusters increase the purity also get improved.This indicates that under number of classes, our method can act as a data mining tool for signal labeling since the human operator only needs to label the representative samples in each cluster, which dramatically increases efficiency. 

\section{Conclusions}
\label{Conclusions}
In this paper, we propose a deep clustering method to automatically classify the radar waveform without any labels. The experimental results show the proposed method can effectively cluster the radar waveform reaching a performance comparable to the supervised counterpart.However, there are also some shortcomings of the proposed method.Firstly, the experiment is conducted under a simulated situation and in the real-world  environment , more class-independent features would involve thus degenerate the performance.Secondly, the pretext model needs several hours of training time and we found most of the time is wasted on data augmentation.So in the future, we hope to experiment on real-world  data and develop augmentations that are fast and effective. 


\vspace{6pt} 



\bibliographystyle{ieeetr}
\bibliography{manuscript}

\begin{thebibliography}{10}

\bibitem{bluntOverviewRadarWaveform2016}
S.~D. Blunt and E.~L. Mokole, ``Overview of radar waveform diversity,'' {\em
  IEEE Aerospace and Electronic Systems Magazine}, vol.~31, pp.~2--42, Nov.
  2016.

\bibitem{wangAutomaticRadarWaveform2017}
C.~Wang, J.~Wang, and X.~Zhang, ``Automatic radar waveform recognition based on
  time-frequency analysis and convolutional neural network,'' in {\em 2017
  {{IEEE International Conference}} on {{Acoustics}}, {{Speech}} and {{Signal
  Processing}} ({{ICASSP}})}, pp.~2437--2441, Mar. 2017.

\bibitem{quRadarSignalIntrapulse2019}
Z.~Qu, W.~Wang, C.~Hou, and C.~Hou, ``Radar signal intra-pulse modulation
  recognition based on convolutional denoising autoencoder and deep
  convolutional neural network,'' {\em IEEE Access}, vol.~7,
  pp.~112339--112347, 2019.

\bibitem{huynh-theAccurateLPIRadar2021}
T.~{Huynh-The}, V.-S. Doan, C.-H. Hua, Q.-V. Pham, T.-V. Nguyen, and D.-S. Kim,
  ``Accurate {{LPI Radar Waveform Recognition With CWD-TFA}} for {{Deep
  Convolutional Network}},'' {\em IEEE Wireless Communications Letters},
  vol.~10, pp.~1638--1642, Aug. 2021.

\bibitem{wuRadarEmitterSignal2020}
B.~Wu, S.~Yuan, P.~Li, Z.~Jing, S.~Huang, and Y.~Zhao, ``Radar {{Emitter Signal
  Recognition Based}} on {{One-Dimensional Convolutional Neural Network}} with
  {{Attention Mechanism}},'' {\em Sensors (Basel, Switzerland)}, vol.~20,
  p.~6350, Nov. 2020.

\bibitem{yuanIntraPulseModulationClassification2021}
S.~Yuan, B.~Wu, and P.~Li, ``Intra-{{Pulse Modulation Classification}} of
  {{Radar Emitter Signals Based}} on a 1-{{D Selective Kernel Convolutional
  Neural Network}},'' {\em Remote Sensing}, vol.~13, p.~2799, Jan. 2021.

\bibitem{weiIntrapulseModulationRadar2020}
S.~Wei, Q.~Qu, H.~Su, M.~Wang, J.~Shi, and X.~Hao, ``Intra-pulse modulation
  radar signal recognition based on {{CLDN}} network,'' {\em IET Radar, Sonar
  \& Navigation}, vol.~14, no.~6, pp.~803--810, 2020.

\bibitem{SignalEyeAISoftware}
``{{SignalEye AI Software}} for {{Automated Signal Classification}} - {{General
  Dynamics}}..''

\bibitem{yuanSemiSupervisedClassificationIntraPulse2022}
S.~Yuan, P.~Li, B.~Wu, X.~Li, and J.~Wang, ``Semi-{{Supervised Classification}}
  for {{Intra-Pulse Modulation}} of {{Radar Emitter Signals Using Convolutional
  Neural Network}},'' {\em Remote Sensing}, vol.~14, p.~2059, Jan. 2022.

\bibitem{liuUnsupervisedRadarSignal2022}
L.~Liu and S.~Xu, ``Unsupervised radar signal recognition based on multi-block
  \textendash{} {{Multi-view Low-Rank Sparse Subspace Clustering}},'' {\em IET
  Radar, Sonar \& Navigation}, vol.~16, no.~3, pp.~542--551, 2022.

\bibitem{devlinBERTPretrainingDeep2019}
J.~Devlin, M.-W. Chang, K.~Lee, and K.~Toutanova, ``{{BERT}}: {{Pre-training}}
  of {{Deep Bidirectional Transformers}} for {{Language Understanding}},'' in
  {\em Proceedings of the 2019 {{Conference}} of the {{North American Chapter}}
  of the {{Association}} for {{Computational Linguistics}}: {{Human Language
  Technologies}}, {{Volume}} 1 ({{Long}} and {{Short Papers}})}, ({Minneapolis,
  Minnesota}), pp.~4171--4186, {Association for Computational Linguistics},
  June 2019.

\bibitem{chenSimpleFrameworkContrastive2020}
T.~Chen, S.~Kornblith, M.~Norouzi, and G.~Hinton, ``A {{Simple Framework}} for
  {{Contrastive Learning}} of {{Visual Representations}},'' {\em
  arXiv:2002.05709 [cs, stat]}, June 2020.

\bibitem{niuSPICESemanticPseudolabeling2022}
C.~Niu, H.~Shan, and G.~Wang, ``{{SPICE}}: {{Semantic Pseudo-labeling}} for
  {{Image Clustering}},'' Jan. 2022.

\bibitem{vangansbekeSCANLearningClassify2020}
W.~Van~Gansbeke, S.~Vandenhende, S.~Georgoulis, M.~Proesmans, and L.~Van~Gool,
  ``{{SCAN}}: {{Learning}} to {{Classify Images}} without {{Labels}},'' {\em
  arXiv:2005.12320 [cs]}, July 2020.

\bibitem{ravikishoreAutomaticIntrapulseModulation2017}
T.~Ravi~Kishore and K.~D. Rao, ``Automatic {{Intrapulse Modulation
  Classification}} of {{Advanced LPI Radar Waveforms}},'' {\em IEEE
  Transactions on Aerospace and Electronic Systems}, vol.~53, pp.~901--914,
  Apr. 2017.

\bibitem{devriesImprovedRegularizationConvolutional2017a}
T.~DeVries and G.~W. Taylor, ``Improved {{Regularization}} of {{Convolutional
  Neural Networks}} with {{Cutout}},'' Nov. 2017.

\bibitem{clarkeStatisticalTheoryMobileradio1968}
R.~H. Clarke, ``A statistical theory of mobile-radio reception,'' {\em The Bell
  System Technical Journal}, vol.~47, pp.~957--1000, July 1968.

\bibitem{limFastAutoAugment2019}
S.~Lim, I.~Kim, T.~Kim, C.~Kim, and S.~Kim, ``Fast {{AutoAugment}},'' in {\em
  Advances in {{Neural Information Processing Systems}}}, vol.~32, {Curran
  Associates, Inc.}, 2019.

\bibitem{caronDeepClusteringUnsupervised2018}
M.~Caron, P.~Bojanowski, A.~Joulin, and M.~Douze, ``Deep {{Clustering}} for
  {{Unsupervised Learning}} of {{Visual Features}},'' in {\em Computer
  {{Vision}} \textendash{} {{ECCV}} 2018} (V.~Ferrari, M.~Hebert,
  C.~Sminchisescu, and Y.~Weiss, eds.), vol.~11218, ({Cham}), pp.~139--156,
  {Springer International Publishing}, 2018.

\bibitem{khoslaSupervisedContrastiveLearning2020}
P.~Khosla, P.~Teterwak, C.~Wang, A.~Sarna, Y.~Tian, P.~Isola, A.~Maschinot,
  C.~Liu, and D.~Krishnan, ``Supervised {{Contrastive Learning}},'' in {\em
  Advances in {{Neural Information Processing Systems}}}, vol.~33,
  pp.~18661--18673, {Curran Associates, Inc.}, 2020.

\bibitem{leePseudoLabelSimpleEfficient2013}
D.-h. Lee, ``Pseudo-{{Label}}: {{The Simple}} and {{Efficient Semi-Supervised
  Learning Method}} for {{Deep Neural Networks}},'' in {\em {{ICML Workshop}}
  on {{Challenges}} in {{Representation Learning}}}, 2013.

\bibitem{sohnFixMatchSimplifyingSemiSupervised2020}
K.~Sohn, D.~Berthelot, N.~Carlini, Z.~Zhang, H.~Zhang, C.~A. Raffel, E.~D.
  Cubuk, A.~Kurakin, and C.-L. Li, ``{{FixMatch}}: {{Simplifying
  Semi-Supervised Learning}} with {{Consistency}} and {{Confidence}},'' in {\em
  Advances in {{Neural Information Processing Systems}}}, vol.~33,
  pp.~596--608, {Curran Associates, Inc.}, 2020.

\bibitem{zhangFlexMatchBoostingSemiSupervised2021}
B.~Zhang, Y.~Wang, W.~Hou, H.~WU, J.~Wang, M.~Okumura, and T.~Shinozaki,
  ``{{FlexMatch}}: {{Boosting Semi-Supervised Learning}} with {{Curriculum
  Pseudo Labeling}},'' in {\em Advances in {{Neural Information Processing
  Systems}}}, vol.~34, pp.~18408--18419, {Curran Associates, Inc.}, 2021.

\bibitem{wangTransferredDeepLearning2019}
Q.~Wang, P.~Du, J.~Yang, G.~Wang, J.~Lei, and C.~Hou, ``Transferred deep
  learning based waveform recognition for cognitive passive radar,'' {\em
  Signal Processing}, vol.~155, pp.~259--267, Feb. 2019.

\bibitem{westDeepArchitecturesModulation2017}
N.~E. West and T.~O'Shea, ``Deep architectures for modulation recognition,'' in
  {\em 2017 {{IEEE International Symposium}} on {{Dynamic Spectrum Access
  Networks}} ({{DySPAN}})}, pp.~1--6, Mar. 2017.

\bibitem{liuTERASelfSupervisedLearning2021}
A.~T. Liu, S.-W. Li, and H.-y. Lee, ``{{TERA}}: {{Self-Supervised Learning}} of
  {{Transformer Encoder Representation}} for {{Speech}},'' {\em IEEE/ACM
  Transactions on Audio, Speech, and Language Processing}, vol.~29,
  pp.~2351--2366, 2021.

\bibitem{vaswaniAttentionAllYou2017}
A.~Vaswani, N.~Shazeer, N.~Parmar, J.~Uszkoreit, L.~Jones, A.~N. Gomez,
  {\L}.~Kaiser, and I.~Polosukhin, ``Attention is {{All}} you {{Need}},'' in
  {\em Advances in {{Neural Information Processing Systems}}}, vol.~30, {Curran
  Associates, Inc.}, 2017.

\bibitem{ioffeBatchNormalizationAccelerating2015}
S.~Ioffe and C.~Szegedy, ``Batch {{Normalization}}: {{Accelerating Deep Network
  Training}} by {{Reducing Internal Covariate Shift}},'' in {\em International
  {{Conference}} on {{Machine Learning}}}, pp.~448--456, {PMLR}, June 2015.

\bibitem{paszkePyTorchImperativeStyle2019}
A.~Paszke, S.~Gross, F.~Massa, A.~Lerer, J.~Bradbury, G.~Chanan, T.~Killeen,
  Z.~Lin, N.~Gimelshein, L.~Antiga, A.~Desmaison, A.~Kopf, E.~Yang, Z.~DeVito,
  M.~Raison, A.~Tejani, S.~Chilamkurthy, B.~Steiner, L.~Fang, J.~Bai, and
  S.~Chintala, ``{{PyTorch}}: {{An Imperative Style}}, {{High-Performance Deep
  Learning Library}},'' in {\em Advances in {{Neural Information Processing
  Systems}}}, vol.~32, {Curran Associates, Inc.}, 2019.

\bibitem{jingAdaptiveFocalLoss2022}
Z.~Jing, P.~Li, B.~Wu, S.~Yuan, and Y.~Chen, ``An {{Adaptive Focal Loss
  Function Based}} on {{Transfer Learning}} for {{Few-Shot Radar Signal
  Intra-Pulse Modulation Classification}},'' {\em Remote Sensing}, vol.~14,
  p.~1950, Jan. 2022.

\bibitem{xieUnsupervisedDeepEmbedding2016}
J.~Xie, R.~Girshick, and A.~Farhadi, ``Unsupervised {{Deep Embedding}} for
  {{Clustering Analysis}},'' in {\em Proceedings of {{The}} 33rd
  {{International Conference}} on {{Machine Learning}}}, pp.~478--487, {PMLR},
  June 2016.

\bibitem{mcconvilleN2DNotToo2021}
R.~McConville, R.~{Santos-Rodr{\'i}guez}, R.~J. Piechocki, and I.~Craddock,
  ``{{N2D}}: ({{Not Too}}) {{Deep Clustering}} via {{Clustering}} the {{Local
  Manifold}} of an {{Autoencoded Embedding}},'' in {\em 2020 25th
  {{International Conference}} on {{Pattern Recognition}} ({{ICPR}})},
  pp.~5145--5152, Jan. 2021.

\bibitem{mcinnesUMAPUniformManifold2020}
L.~McInnes, J.~Healy, and J.~Melville, ``{{UMAP}}: {{Uniform Manifold
  Approximation}} and {{Projection}} for {{Dimension Reduction}},'' Sept. 2020.

\bibitem{chenInfoGANInterpretableRepresentation2016}
X.~Chen, Y.~Duan, R.~Houthooft, J.~Schulman, I.~Sutskever, and P.~Abbeel,
  ``{{InfoGAN}}: {{Interpretable Representation Learning}} by {{Information
  Maximizing Generative Adversarial Nets}},'' in {\em Advances in {{Neural
  Information Processing Systems}}}, vol.~29, {Curran Associates, Inc.}, 2016.

\bibitem{gongUnsupervisedSpecificEmitter2020}
J.~Gong, X.~Xu, and Y.~Lei, ``Unsupervised {{Specific Emitter Identification
  Method Using Radio-Frequency Fingerprint Embedded InfoGAN}},'' {\em IEEE
  Transactions on Information Forensics and Security}, vol.~15, pp.~2898--2913,
  2020.

\end{thebibliography}
\end{document}